\documentclass[a4paper,10pt]{article}
\usepackage{graphicx}
\usepackage{dcolumn}
\usepackage{amsmath}
\usepackage{amssymb}
\usepackage{bm}
\usepackage{epsf}
\usepackage{rotate}
\usepackage{color}
\usepackage{times}
\usepackage{hyperref}

\newcommand{\dx}{{\rm d}x}
\renewcommand{\i}{{\rm i}}
\newcommand{\e}{{\rm e}}
\newcommand{\ie}{{\it i.e.~}}
\renewcommand{\d}{{\rm d}}

\newcommand{\lab}[1]{}
\newcommand{\Or}{{\cal O}}
\newcommand{\be}{\begin{equation}}
\newcommand{\ee}{\end{equation}}
\newcommand{\mycaption}[1]{\caption{\footnotesize #1}}

\def\PutFigure#1{
    \noindent\hfil\includegraphics[scale=.4,angle=-90]{#1}
}

\begin{document}
\small
{\footnotesize
\title{Negative radiation pressure in case of two interacting fields}

\author{Tomasz Roma\'nczukiewicz\\{\tt trom@th.if.uj.edu.pl}\\Institute of Physics,\\
       Jagiellonian University, Reymonta 4, Cracow, Poland}

\maketitle

\begin{abstract}
We study a simple toy model, which although probably does not have any direct physical applications,
can serve as a nice pedagogical example for explanation strange phenomenon of {\it negative radiation pressure} and can also give some insight for understanding the interaction between radiation and a vortex. The model discussed is a classical field theory of two interacting scalar fields in 1+1 d. Without coupling one of the fields is governed by the ordinary $\phi^4$ equation and the second one obeys the Klein-Gordon equation. The possibility of existence of the {\it negative radiation pressure} with a respect to the mass ratio of those two fields is discussed. 
\end{abstract}
}


\section{Intoduction}
In modern physics topological defects play a very special role. Because of huge variety of topological defects they found applications in many branches of physics starting from glueballs and fluxtubes in QCD through vortices in quantum liquids and superconductivity, ferromagnets to cosmological strings and domain walls.
They owe their popularity to their stability due to nontrivial asymptotic behaviour. In many processes they can be treated as ordinary pointlike particles (kinks, monopoles), strings or membranes. However they have also many properties when their internal complex structure becomes dominant showing that topological defects can be very different from particles or other simple objects. 

One of the phenomenon in which, on the first sight, topological defects behave like ordinary particles is radiation pressure. When a kink is hit with scalar radiation it reflects some of this scalar field. 
This reflection pushes the kink. This effect can be effectively described as radiation pressure very similar in its nature to radiation pressure known from electrodynamics. 
However there is a large group of topological defects which behave in a completely different manner revealing the negative radiation pressure (NRP) like kinks in $\phi^4$ \cite{trom0,forgacs} and vortices in Goldston's and abelian Higgs models.

The NRP was first discovered in $\phi^4$ model. When the plane wave hit the kink it was accelerating towards the source of radiation. This effect was confirmed using both numerical simulation and perturbation theory.
In the first order the kink is transparent to the wave due to the so called reflectionless potential. 
Therefore our usual intuition can be missleading.
The further corrections showed that after the transition the wave carried the same amount of energy but more momentum. The surplus of momentum forces the kink to accelerate towards the source of radiation.
The force is proportional to $A^4$. In other models which do not have kinks with reflectionless potentials the solitons behave as one could expect. They are pushed by radiation and the leading term in the force is proportional to $A^2$.

Here we propose a different example of the negative radiation pressure. We have decided to discuss a strange model because of a few reasons. The appearance of the NRP in this model has a different mechanism. The leading term in the force is proportional to $A^2$ and therefore it suggests that it comes in the first order of perturbation scheme. Moreover the system possesses another parameter (a mass of Klein-Gordon field) which one can change and find two different regimes: a negative radiation pressure and the positive one. 
Our preliminary numerical calculations suggest that similar mechanism is responsible for the  NRP in case of vortices. The NRP is proportional again  to $A^2$. Moreover in the abelian Higgs model there is also a parameter which can be triggered. It is usually used to distinguish between superconductors type I and type II. 
This parameter possesses also a different value which distinguishes between vortices exhibiting NRP and PRP. 

The model discused in the present paper has very similar properties of the radiation pressure as Goldstone's model. The adventage of the model discussed is its simplicity.

The paper is constructed in the following way. First we give a motivation as to why we have chosen this particular model. Next section is a description of the model itself and some spacial solutions existing in this model. In the following section we analyse small perturbations around topological defect using standard perturbation procedure both in coupling constant and wave amplitude. Within this perturbation scheme we construct a wave travelling from rhs of the kink. We derive an analytic form for the force exerted on the kink by the wave.
The next section is devoted to numerical results where we compare our perturbation scheme results with numerical solution to the full partial differential equation.
Finally we summarise the results and discuss possible applications and extensions.

In the Appendices we give some more detailed calculations. The last Appendix is a short description of Goldstone's and Higgs modeles pointing out the similarities to the model presented in this paper.

\section{Motivation}
As we mentioned before the negative radiation pressure was first observed numerically and then explained analytically in case of one-dimensional kinks in $\phi^4$ model. The key to this unusual phenomenon is transparency of the kink in linear regime. The second order calculation showed that a wave with twice the frequency is created and that wave carries more momentum than the incident wave creating the surplus of momentum behind the kink. Because of the energy and momentum conservation laws the surplus of momentum must be compensate with kink's motion. The kink starts moving towards the source of radiation. This is purely nonlinear effect.

The feature of reflectionless in linear regime is rather rare and does not necessarily lead to the negative radiation pressure. As a counter-example in paper \cite{forgacs} we have discussed another often studied model \ie sine-Gordon. Kinks in this model are also transparent but they are transparent exactly to all orders in perturbation series and hence experience no radiation pressure at all. 

Fortunately the phenomenon of NRP turned out to be quite robust with respect to perturbation of the $\phi^4$ potential. Although the reflectionless was lost the nonlinear effects were still visible and above a certain critical amplitude the NRP was present even for quite large deformation of the potential.

As we stated in the introduction the NRP is visible not only for kinks in 1+1 d but also for Goldstone's and abelian Higgs model vortices. Unfortunately the analysis of the negative radiation pressure is difficult for those models both analytically and numerically and will be presented in separate papers. We believe that in case of vortices the negative radiation pressure has a different nature than in $\phi^4$ model.
Our preliminary numerical analysis shows that in case of vortices the force exerted by travelling wave is proportional to $A^2$ on the contrary to $A^4$ in $\phi^4$ model. Moreover the vortices are clearly not reflectionless and hence a different mechanism must take place.

In the present paper we propose a model exhibiting kinks undergoing negative radiation pressure proportional to $A^2$. We explain the mechanism which stands behind this process which is different than the one in $\phi^4$ model. The model is quite simple and probably does not have any direct physical application but some of the features seem to be generic and are not blurred with technical difficulties.

When the negative radiation pressure is present there always must be some kind of mechanism creating a surplus of momentum behind a defect. In $\phi^4$ case the momentum to energy ratio was $k(\omega)/\omega=\sqrt{\omega^2-4}/\omega$. The wave with twice the frequency has momentum to energy ratio equal to $k(2\omega)/2\omega=\sqrt{\omega^2-1}/\omega>k(\omega)/\omega$. Therefore a wave with twice the frequency created due to the nonlinearities carries more momentum than the incoming wave.

This is not the only possible way to create a surplus of momentum. Let us consider two scalar fields with masses $m_1$ and $m_2$ scattered on some object (\ie topological defect). Suppose that only a wave with mass $m_1$ comes from one side of the defect. During scattering process the wave is converted into a wave with mass $m_2$. The momentum to energy ratio is equal to $\sqrt{\omega^2-m_i}/\omega$ and therefore if $m_1>m_2$ we have again a surplus of momentum behind the object which could undergo the negative radiation pressure. Similar process is possible for a single scalar field with degenerate vacua when the excitations around those vacua have different masses. 

\section{Model}
Let us now propose a simple model of two interacting fields with two different masses. We want the model to exhibit topological defects. A natural candidate would be a $\phi^4$ model field coupled with some other field\footnote{Other possible models are discussed in section \ref{sec:other}}:
\begin{equation}
{\cal L}=\frac{1}{2}\left(\phi_t^2-\phi_x^2\right)+\frac{1}{2}\left(\psi_t^2-\psi_x^2\right)-
\frac{1}{2}\biggl((\phi^2-1)+\kappa\psi\biggr)^2-\frac{1}{2}(m^2-\kappa^2)\psi^2
\end{equation}
or equivalently
\begin{equation}\label{lag2}
{\cal L}=\frac{1}{2}\biggl(\phi_t^2-\phi_x^2-(\phi^2-1)^2\biggr)+\frac{1}{2}\left(\psi_t^2-\psi_x^2-m^2\psi^2\right)-
\kappa(\phi^2-1)\psi,
\end{equation}
where $m$ is a mass of $\psi$ field and $\kappa$ is a coupling constant. Note that when $\kappa^2=m^2$ the vacuum manifold satisfies equation $\psi = (1-\phi^2)/\kappa$ and no topological defect is possible. Otherwise the vacuum consists of merely two disconnected points $(\phi=\pm 1, \psi=0)$ revealing $\mathbb{Z}_2$ symmetry and domain walls (in general) or kinks (in one-dimensional models) are possible. 

\begin{figure}
\PutFigure{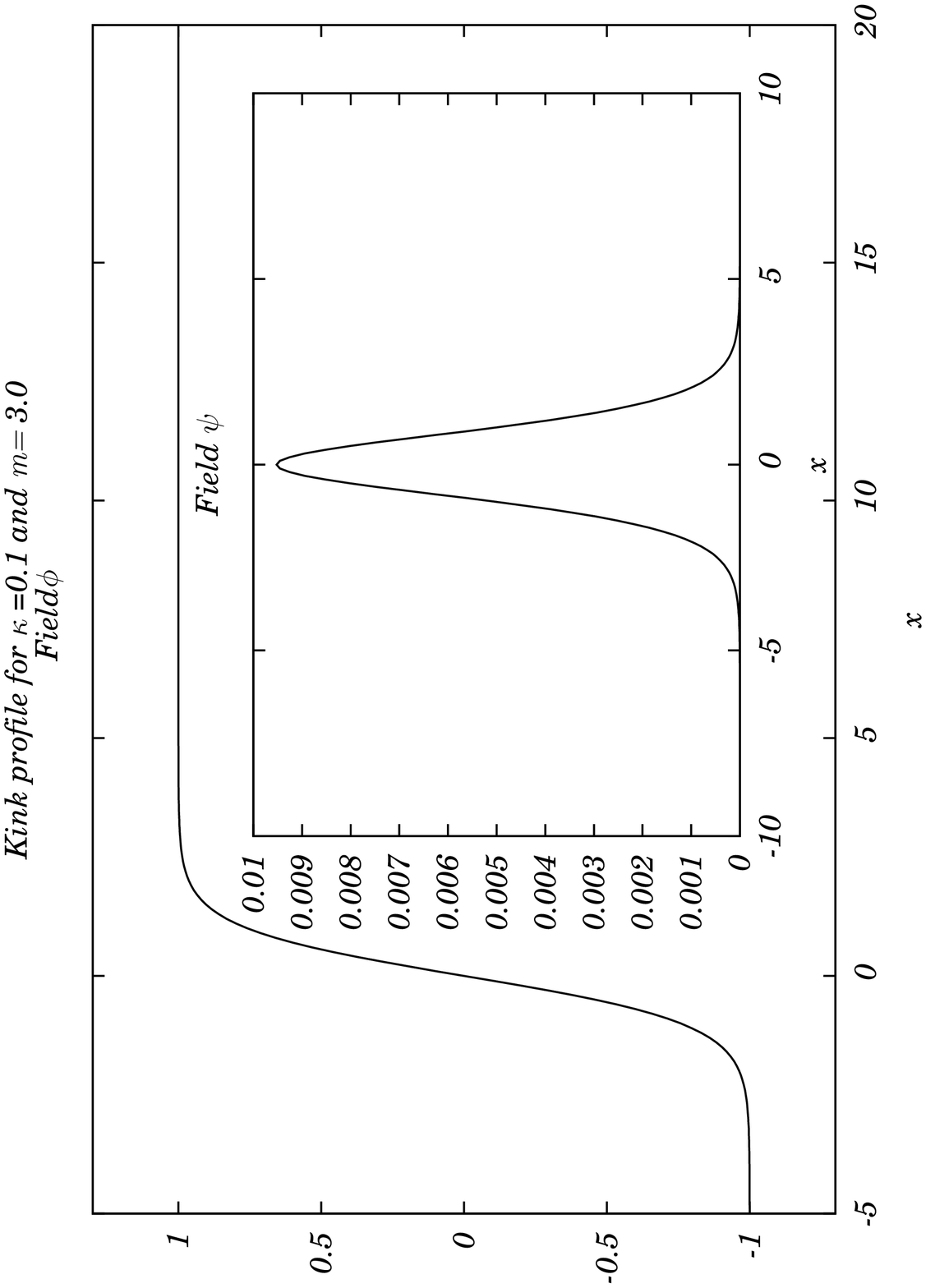}
\mycaption{\label{figure1}Static solutions for the kink for $\kappa=0.1$ and $m=3.0$}
\end{figure}

The equations of motion are:
\begin{equation}\label{eq:PDE}
 \begin{cases}
  \phi_{tt}-\phi_{xx}+2\phi(\phi^2-1)+2\kappa\phi\psi&=0,\\
  \psi_{tt}-\psi_{xx}+m^2\psi+\kappa(\phi^2-1)&=0.
 \end{cases}
\end{equation}
Unfortunately we do not know even nontrivial time independent solution -- the kinks in analytical form.
Numerically calculated profiles are depicted in Figure \ref{figure1}. For small values of $\kappa$ the $\phi$ field is very similar to $\tanh x$ as in ordinary $\phi^4$ model. The $\psi$ field is a gaussian-shape function with its height proportional approximetly to $\kappa$.
Large $x$ value asymptotic approximation gives  $\phi\approx1-a\e^{-nx}$ and $\psi\sim\e^{-nx}$ with $n$ being the smaller value out of:
\begin{equation}
n = \sqrt{\frac{m^2+4\pm\sqrt{(m^2-4)^2+16\kappa^2}}{2}}.
\end{equation}

\section{Radiation pressure}
\subsection{Planar wave}
Let us consider a process when a kink is hit with a $\psi$-field wave with frequency $\omega$ coming from $+\infty$.
In order to calculate the leading term to the radiation pressure we do not need to know the full kink solution. All we need is the the perturbation series in coupling constant $\kappa$: $\phi=\phi^{(0)}+\kappa\phi^{(1)}+\kappa^2\phi^{(2)}+\cdots$ and $\psi=\psi^{(0)}+\kappa\psi^{(1)}+\kappa^2\psi^{(2)}+\cdots$.

In $\Or(\kappa^0)$ the equations take the form:

\begin{subequations}

 \begin{align}\label{eq:order0a}
  \phi^{(0)}_{tt}-\phi^{(0)}_{xx}+2\phi^{(0)}\left({\phi^{(0)}}^2-1\right)&=0.\\
 \label{eq:order0b}
  \psi^{(0)}_{tt}-\psi^{(0)}_{xx}+m^2\psi^{(0)}&=0.
 \end{align}
\end{subequations}

For static kink solution from these equation we obtain $\phi^{(0)}=\tanh x$ and $\psi^{(0)}=0$.
In order to study a scattering process instead of $\psi^{(0)}=0$ we take $\psi^{(0)}=A\cos(k x+\omega t)$ where $A$ is an amplitude of the wave and $k=\sqrt{\omega^2-m^2}$ is a wave number. 
The first order equations are:

\begin{subequations}
 \begin{align}\label{eq:order1a}
  \phi^{(1)}_{tt}-\phi^{(1)}_{xx}+2\left(3{\phi^{(0)}}^2-1\right)\phi^{(1)}+2\phi^{(0)}\psi^{(0)}&=0.\\
 \label{eq:order1b}
  \psi^{(1)}_{tt}-\psi^{(1)}_{xx}+m^2\psi^{(1)}+\left({\phi^{(0)}}^2-1\right)&=0.
 \end{align}
\end{subequations}
The equation (\ref{eq:order1b}) has only static inhomogeneous part so it gives only correction to the static solution of the kink. Since we know solutions to the homogeneous part (for $\psi^{(1)}_t=0$) we can express the solution of inhomogeneous equation in terms of Green's function: 
\begin{equation}
 \psi^{(1)} = \frac{1}{2m}\left(\e^{-mx}\int_{-\infty}^x\!\!\dx'\;\frac{\e^{mx'}}{\cosh^2x'} + \e^{mx}\int^{\infty}_x\!\!\dx'\;\frac{\e^{-mx'}}{\cosh^2x'}\right).
\end{equation}
These integrands can be calculated using hypergeometric functions.

Since $\psi^{(0)}=\frac{1}{2}\e^{\i \omega t+kx}+c.c.$ the solution to the equation (\ref{eq:order1a}) can be sought in the fallowing form:
\begin{equation}
 \phi^{(1)} = \frac{1}{2}\e^{\i \omega t}\xi_+(x)+\frac{1}{2}\e^{-\i \omega t}\xi_-(x)
\end{equation}
where $\xi_+$ is a solution to
\begin{equation}\label{eq:inhomo}
 \left(\frac{\d^2}{\dx^2}+q^2+\frac{6}{\cosh^2 x}\right)\xi_+=2A\e^{\i kx}\tanh x,
\end{equation}
with $q^2=\omega^2-4$ and $\xi_-=\xi_+^*$. Fortunately we already know the solution $\eta_q$ to the homogeneous part of the above equation
 (see \cite{forgacs})\footnote{Note that there is no reflection part proportional to $\e^{-\i qx}$} 
 \begin{equation}
  \eta_q(x) = \frac{3\tanh^2 x-1-q^2-3\i q\tanh x}{\sqrt{(q^2+1)(q^2+4)}}\e^{\i qx}
 \end{equation}
and again we can use the Green function technique to construct the solution:
\begin{equation}
 \xi_+(x) = -\frac{\eta_{-q}}{W}\int_{-\infty}^x\!\!\dx'\;\eta_q(x')f(x')-
\frac{\eta_{q}}{W}\int_{x}^\infty\!\!\dx'\;\eta_{-q}(x')f(x'),
\end{equation}
where $W=-2iq$ is Wronskian and $f(x)=A\e^{\i kx}\tanh x$ is the {\it rhs} of the equation (\ref{eq:inhomo}).
As for all scattering processes we are interested only in the asymptotic form of this solution. Therefore it is convenient to rewrite the above solution for large values of $x$ in a little different form:
\begin{equation}\label{eq:intrewrited}
 \xi=-\frac{\eta_{-q}}{W}\int_{-\infty}^\infty\!\!\dx'\;\eta_q(x')f(x')+
\frac{1}{W}\int_{x}^\infty\!\!\dx'\;f(x')\left(\eta_{-q}(x)\eta_q(x')-\eta_{q}(x)\eta_{-q}(x')\right).
\end{equation}
To calculate the second integral we can use the asymptotic forms both of $\eta_q(x)$
\begin{equation}
 \eta_q(x\rightarrow\infty)\approx\frac{2-q^2-3\i q}{\sqrt{(q^2+1)(q^2+4)}}\e^{\i qx}
\end{equation}
and $f(x)=2A\e^{\i kx}$ to obtain:
\begin{equation}
\begin{split}
 I\equiv\frac{1}{W}\int_{x}^\infty\!\!\dx'\;f(x')\left(\eta_{-q}(x)\eta_q(x')-\eta_{q}(x)\eta_{-q}(x')\right)
=\\
=\frac{2}{W}\int_{x}^\infty\!\!\dx'\biggl(\exp\bigl(\i  ( qx'-qx+kx')\bigr)-
                                          \exp\bigl((\i (-qx'+qx+kx')\bigr)\biggr).
\end{split}
\end{equation}
With help  of identity
\begin{equation}
 \int_x^\infty\!\!\dx'\;\e^{\i ax'}=\pi\delta(a)+\frac{\i}{a}\e^{\i ax}
\end{equation}
we obtain
\begin{equation}
 I = 2\frac{\e^{\i kx}}{k^2-q^2}+\frac{2\pi\e^{\i kx}}{W}\left(\delta(k+q)-\delta(k-q)\right).
\end{equation}
Because only when $m=2$ (when model does not exhibit topological defects) the $\delta$s give contribution, $I$ represents only the inhomogeneous part of the solution. Note that the same solution is present for $x\rightarrow-\infty$.

The first integral of equation (\ref{eq:intrewrited}) can be calculated via residua method. The solution for large values of $x$ can therefore be written as:

\begin{equation}
 \xi_+(x\rightarrow+\infty)=R(q,k)\eta_{-q}(x)-\frac{\e^{\i kx}}{k^2-q^2}.
\end{equation}
where
\begin{equation}
 R(q,k)=\frac{\pi\left(3k^2-q^2-4\right)}{2q\sqrt{(q^2+1)(q^2+4)}
	\sinh\left(\frac{q+k}{2}\pi\right)}
\end{equation}
is reflection coefficient. For $x\rightarrow-\infty$ similar procedure leads to the following solution
\begin{equation}
 \xi_+(x\rightarrow-\infty)=T(q,k)\eta{-q}(x)+\frac{\e^{\i kx}}{k^2-q^2}.
\end{equation}
where $T(q,k)=R(-q,k)$ is the transition coefficient. One can see that for $q=k\Leftrightarrow m=2$ the transition coefficient becomes infinite. Of course this means that our perturbation scheme fails in the vicinity of this point. \\
Another important observation is that for $q^2=3k^2-4\Leftrightarrow\omega^2=\frac{3m^2}{2}$ the nominator vanishes and the kink becomes again. This should be clearly seen in numerical simulations. 
\subsection{The force}

Having the asymptotic form of solution representing travelling wave 
\begin{subequations}
\begin{equation}\label{eq:travelling}
 \phi(x,t) = \begin{cases}
         \frac{1}{2}A\kappa\e^{\i\omega t}\left(R(q,k)\eta_{-q}(x)-\frac{\e^{\i kx}}{k^2-q^2}\right)+c.c.
	 \;\;\; &\textrm{for}\;\;x-\rightarrow+\infty\\
	\frac{1}{2}A\kappa\e^{\i\omega t}\left(T(q,k)\eta_{q}(x)+\frac{\e^{\i kx}}{k^2-q^2}\right)+c.c.
	 \;\;\; &\textrm{for}\;\;x-\rightarrow-\infty
        \end{cases}
\end{equation}
\begin{equation}\label{eq:travelling2}
 \psi(x,t) = \begin{cases}
         A\cos(\omega t + kx)
	 \;\;\; &\textrm{for}\;\;x\rightarrow+\infty\\
	Ab\cos(\omega t+kx)
	 \;\;\; &\textrm{for}\;\;x\rightarrow-\infty
        \end{cases}
\end{equation}
\end{subequations}
we can calculate the force which is exerted on the kink by this wave. In the second equation we changed the amplitude from $A$ to $Ab$. Actually as we will shortly show $b=1+\Or(\kappa ^2)$ so this correction is not visible at  this point of perturbation series but it is necessary to fulfil the energy conservation law. If the kink is initially not moving the rate of energy flowing into a large box containing the kink is given by (for more details see Appendix \ref{sec:noether})
\begin{equation}
\left.\langle\dot E\rangle_T = \langle\phi_t\phi_x+\psi_t\psi_x\rangle_T\right|_{-L}^L,
\end{equation}
were we average over a period. After substitution we obtain:
\begin{equation}\label{eq:energy}
 \left\langle\frac{\d E}{\d t}\right\rangle_T = \frac{1}{2}\omega A^2\left(-q\kappa^2R^2+k-kb^2-q\kappa^2T^2\right)
\end{equation}
and since energy must be conserved ($\langle\dot E\rangle=0$) we obtain
\begin{equation}\label{eq:b}
 b^2=1-\kappa^2\frac{q}{k}\left(R^2+T^2\right).
\end{equation}
note that our scheme is obviously invalid when $b^2<0$ \ie
\begin{equation}
   R^2+T^2>\frac{k}{\kappa^2 q}.
\end{equation} 
This correction to the amplitude of $\psi$ wave would be obtained after calculating next step in our perturbation series but since the energy must be conserved we obtained this result as a consistency condition. (Similar result was discused in more details in \cite{forgacs}.)

The force exerted on the kink can be calculated using the momentum conservation law(for more details see Appendix \ref{sec:noether}):
\begin{equation}
 F = \left\langle\frac{\d P}{\d t}\right\rangle_T = \frac{1}{2}A^2\left(-k^2-\kappa^2R^2q^2+b^2k^2+\kappa^2T^2\right)
\end{equation}
or using (\ref{eq:b}):
\begin{equation}\label{eq:force}
 F = \frac{1}{2}A^2q\left(T^2(q-k)-R^2(q+k)\right).
\end{equation}
The above force is proportional to $A^2$ (contrary to $A^4$ in $\phi^4$ model \cite{forgacs}).
Moreover it can both be positive (kink accelerates towards the source of radiation -- negative radiation pressure) and negative (the kink accelerates with the wave due to the positive radiation pressure). Note that when $q<k$ \ie $m<2$ the force is always negative (or at most zero) whatever the coefficients $R$ and $T$ are.
When $m>2$ the direction in which the kink accelerate can be determined only after substitution the values of $R$ and $T$. The negative radiation pressure appears when
\begin{equation}
 T^2>\frac{q+k}{q-k}R^2.
\end{equation}
This inequality can be rewritten (when $R\neq 0$) as 

\begin{equation}
   \frac{\sinh\frac{q-k}{2}\pi}{\sinh\frac{q+k}{2}\pi} > \frac{q+k}{q-k}
\end{equation} 
which is true for all $q>k>0$. The force vanishes only when $R=0$ that is when $\omega^2=\frac{3m^2}{2}$.

Another important conclusion follows from our considerations which distinguishes the me\-cha\-nizm presented in this paper from the NRP in $\phi^4$. The force in our model comes directly from the first order calculations in amplitude of the wave. We do not need therefore to limit our consideration to monochromatic waves.
Any $\psi$-wave can be expressed as a wave packet:
\begin{equation}
   \psi(x,t)=\int\!\!\d k\;\; A(k)\e^{\i(kx+\omega(k)t)}+c.c.
\end{equation} 
The averaged force exerted by the packet is therefore:
\begin{equation}
{\cal F} = \frac{1}{2}\kappa^2\int\!\!\d k\;\; A^2(k) \left(T^2(k)(q(k)-k)-R^2(k)(q(k)+k)\right).
\end{equation} 
Because when $m>2$ the force is not negative for all frequencies and the kink would always undergo the NRP. In case when $m<2$ the kink would be pushed by the packet.

\begin{figure}
\PutFigure{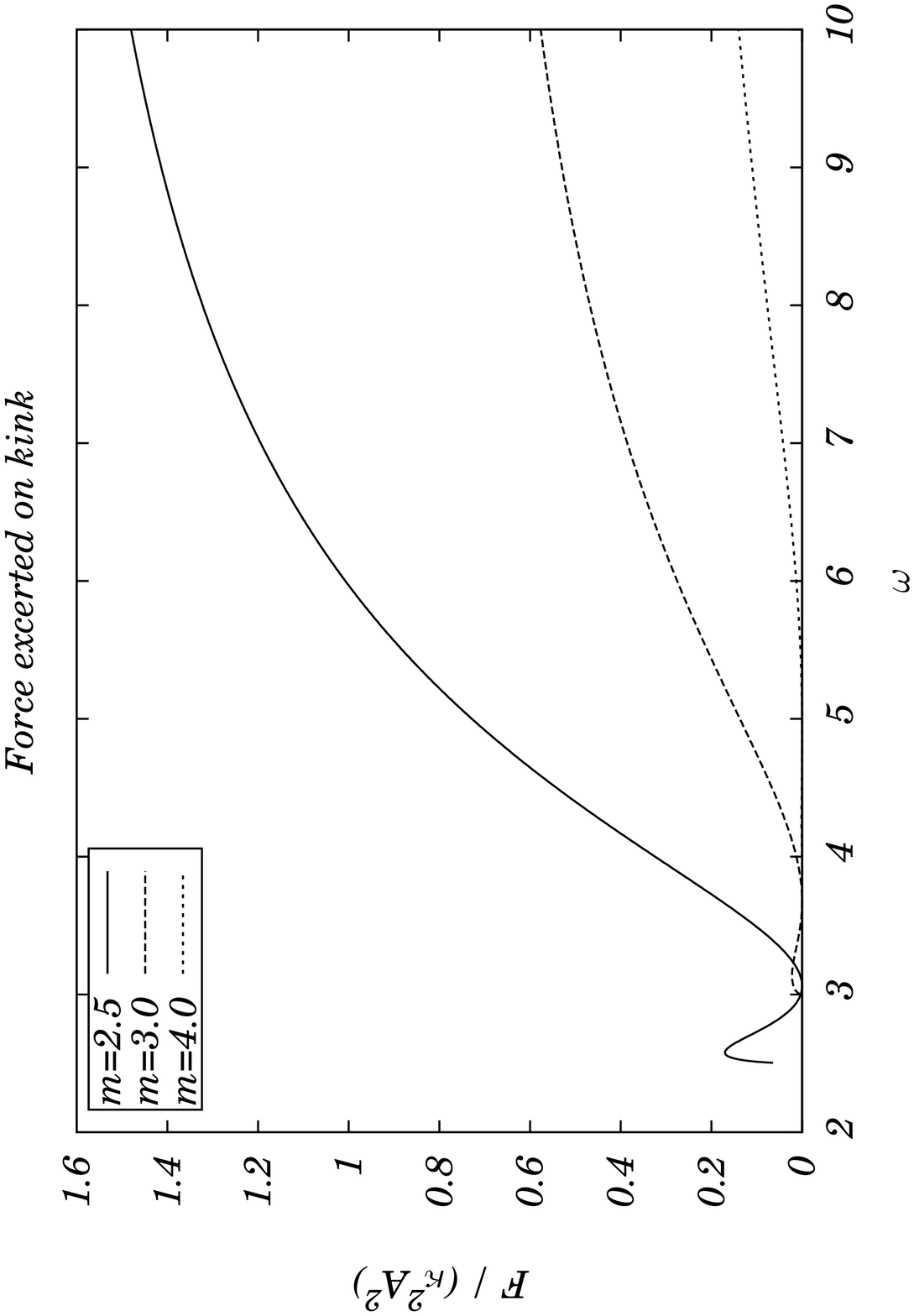}
\mycaption{\label{f2} Force exerted on a kink (divided by $\kappa^2A^2$) for three different values of $m$.}
\end{figure}

\section{Numerical calculations}
As we have stated before, even in this simple model we do not now the exact analytical form of static solutions. Therefore they need to be obtained numerically. We have done this using collocation Chebyshev spectral method in variable $s=\tanh x$. In this way we obtained the solutions $\phi_0(x)$ and $\psi_0(x)$ depicted in the Figure \ref{figure1}.
\

The next step was to find the initial condition representing travelling $\psi$ wave. Obtaining a solution satisfying all assumptions of the travelling wave is rather cumbersome. Therefore we have decided to use only an approximation:
\begin{equation}
   \phi(x,t=0) = \phi_0(x),\;\;\dot  \phi(x,t=0)=0,
\end{equation} 
and
\begin{equation}
   \psi(x,t=0) = \psi_0(x)+A\cos(kx),\;\;\dot  \psi(x,t=0)=-A\omega\sin(kx),
\end{equation} 
Disadvantage of the above initial data is that at the very beginning the kink has a no-zero initial velocity. 
This is due to the fact that the above initial conditions are combination of many eigenstates of the linearised equation around the kink, including translational mode. 
However this type of conditions show how roboust and generic our considerations are.

\begin{figure}
\PutFigure{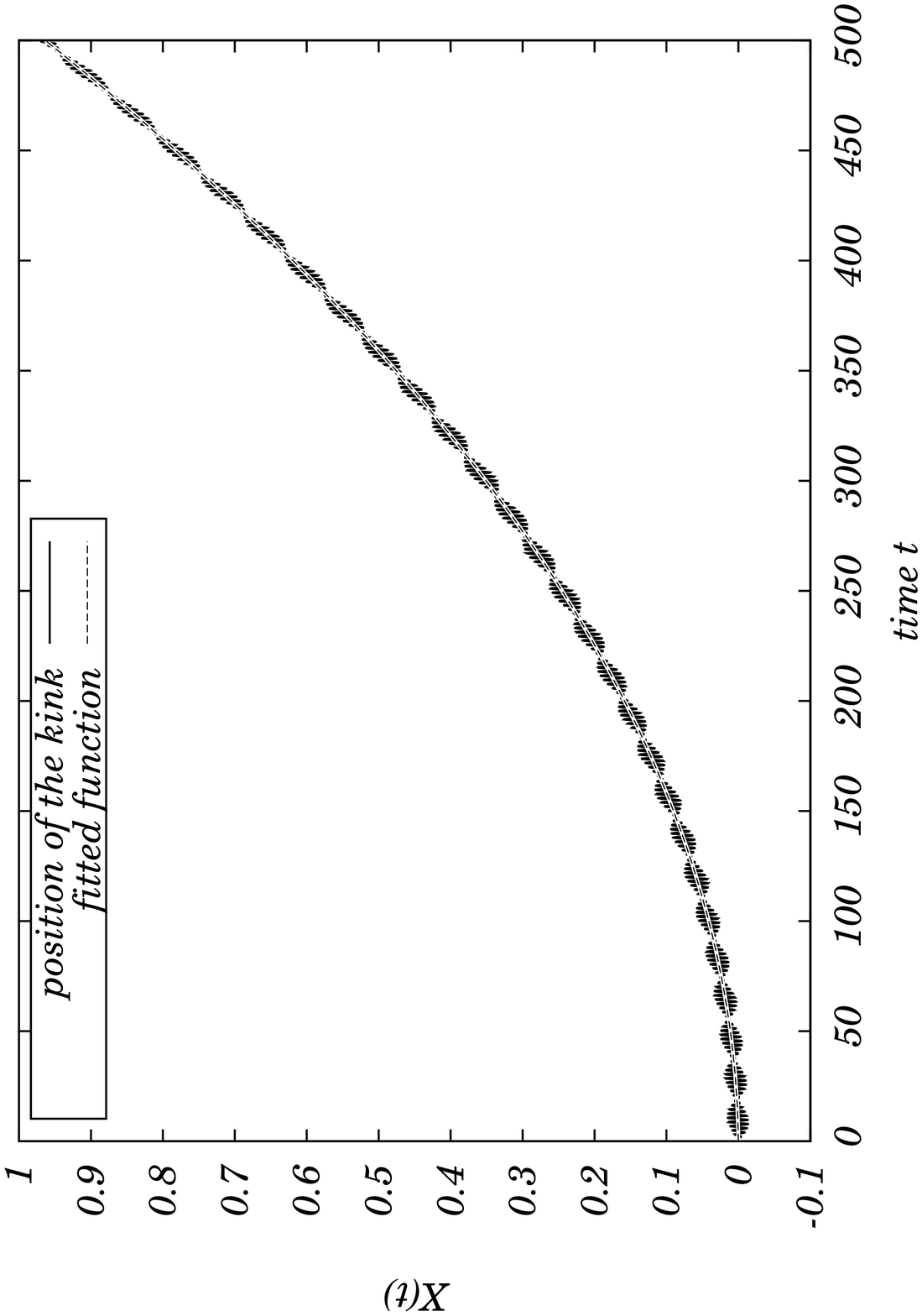}
\mycaption{\label{f3} Trajectory of the kink along with fitted parabola $X(t)=\frac{1}{2}at^2+vt+c$ for $\kappa=0.1$, $m=2.5$, $A=0.1$ and $\omega=3.5$. The fitted values: $a = (7.57898\pm 0.00069)\cdot10^{-6} $, $b= (3.23\pm0.18)\cdot10^{-5}$  and $c = (1.1\pm 1.9)\cdot 10^{-4}$.}
\end{figure}

We solved our PDE equation (\ref{eq:PDE}) using five-point descritization of the second spatial derivative and integrating the obtained system of ODEs with 4th order Runge-Kutta method. We have approximated the position of the kink with the topological zero of the $\phi$ filed. The position of a static kink is well defined and it coincides with the topological zero. In case when the kink is exposed to the wave its position is not well defined but it cannot be far away from topological zero. During its motion the kink recedes from its initial position significantly and those subtleties can be neglected. The example trajectory of the kink is presented in the Figure \ref{f3}. The path of the kink is very similar to parabola (as expected) and we can fit it with a function in the form $X(t)=\frac{1}{2}at^2+vt+c$ obtaining its initial velocity $v$ and acceleration $a$.

\begin{figure}
\PutFigure{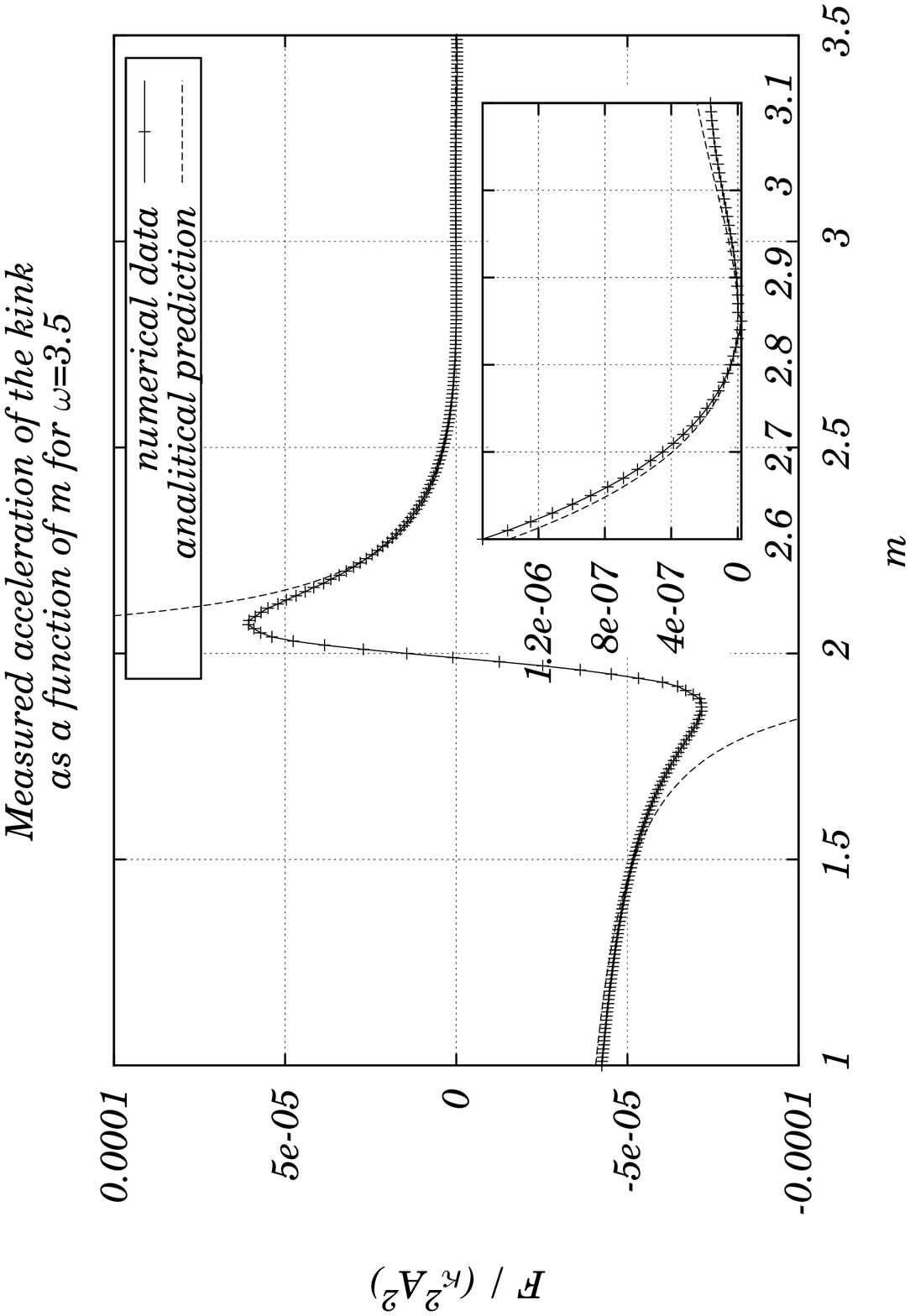}
\mycaption{\label{f4} Fitted acceleration compared with the analytical prediction as a function of $m$. $\omega=3.5$, $\kappa=0.1$, $A=0.1$.}
\end{figure}

\begin{figure}
\PutFigure{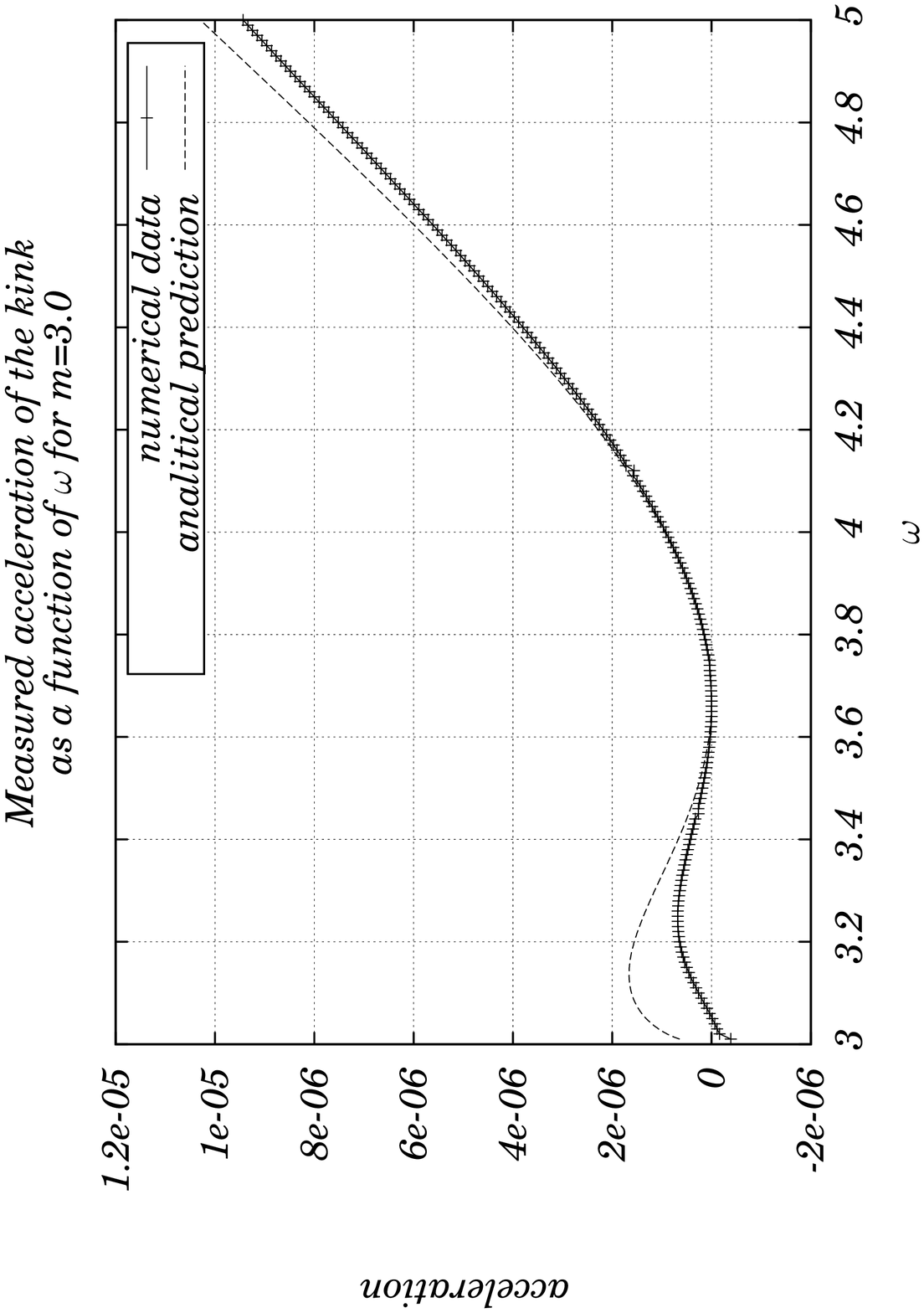}
\mycaption{\label{f5} Fitted acceleration compared with the analytical prediction as a function of $\omega$. $m=3.0$, $\kappa=0.1$, $A=0.1$.}
\end{figure}

In the Figures \ref{f4} and \ref{f5} we have depicted the measured acceleration as a function of mass parameter $m$ and frequency $\omega$ along with theoretically predicted functions (force from equation (\ref{eq:force}) divided by kink mass for $\kappa=0$: $M_k=\frac{4}{3}$). 
Figure \ref{f4} shows that analytical predictions are quite precise (outside the shaded region -- vicinity of $m=2$, when our perturbation scheme fails).
Note that the force disappears (kink is transparent to the radiation) when $2\omega^2=3m^2$. 
This can be seen in both Figures. 
Our analytical prediction differs significantly from numerically obtained acceleration also for small frequencies (Figure \ref{f5}). 
This can be due to the fact that our approximated initial conditions differ significantly from the travelling (eigenfunction) wave, especially for small frequencies.

Except this low frequency limit and vicinity of $m=2$ our theory conicides with numerical simulation with average error of a few percents which is satisfying having in mind all the assumptions and simplifications we used.

\begin{figure}
\PutFigure{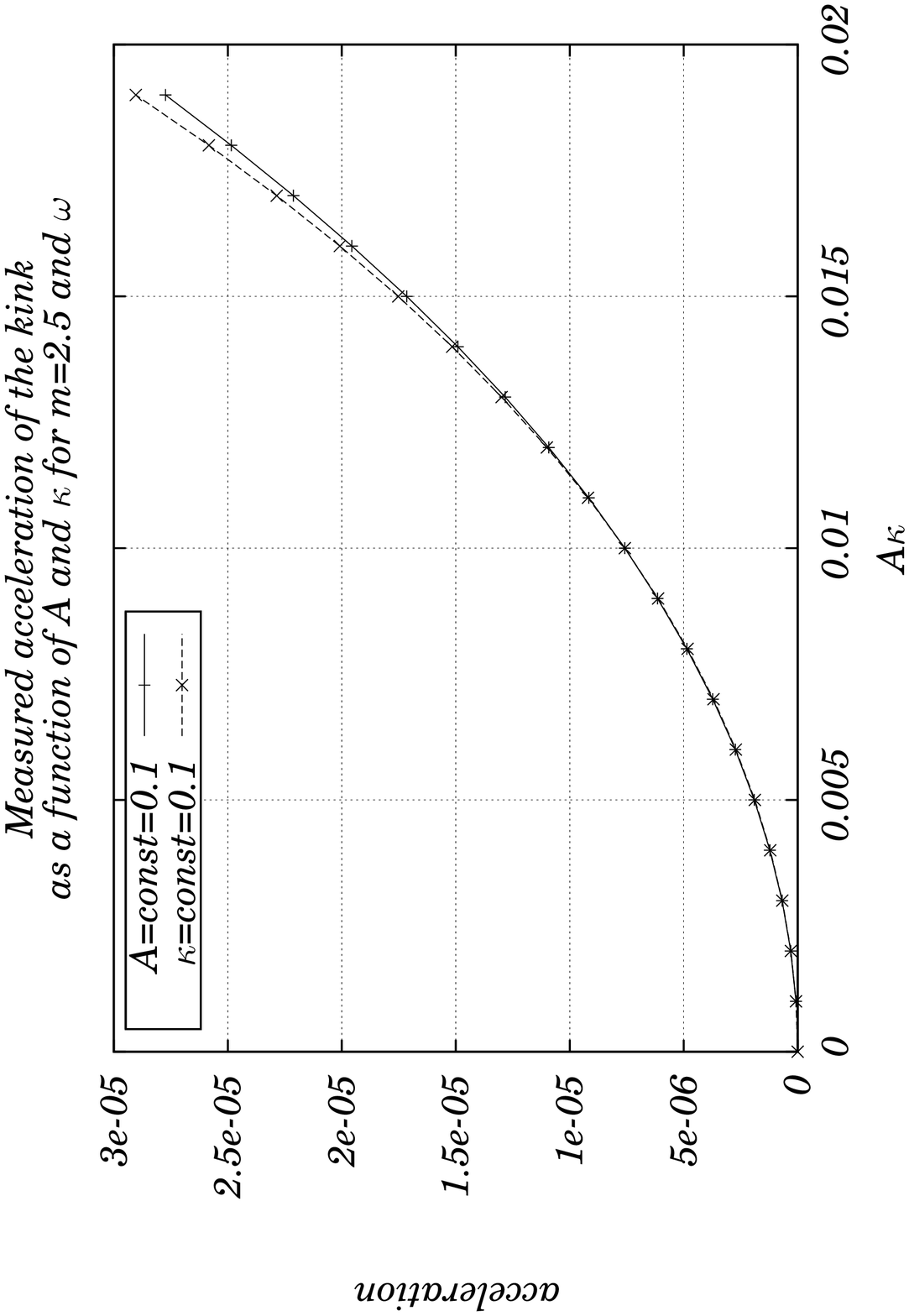}
\mycaption{\label{f6} Fitted acceleration  as a function of $A\kappa$. $\omega=3.5$ and $m=2.5$. One line is for constant $A=0.1$ and the second one is for $\kappa=0.1$}
\end{figure}

The last Figure shows how acceleration grows with amplitude $A$ and a coupling constant $\kappa$. From our theoretical prediction we expect that acceleration should be proportional to $A^2\kappa^2$. In the Figure \ref{f6} we have plotted the acceleration as a function of $A\kappa$. There are two lines: one for constant $A=0.1$ and the second one for constant $\kappa=0.1$. Those functions coincide very well especially when both $\kappa$ and $A$ are small. We have fitted the exponent for both functions in the form $a=\alpha\e^{\beta A\kappa}$. For constant $\kappa$ we obtained $\beta_A = 2.0888 \pm 0.0075$ and for constant $A$: $\beta_\kappa=2.0196 \pm 0.0025$. This indicates that indeed the acceleration is proportional to $A^2\kappa^2$.

\subsection{Other models}\label{sec:other}
\subsubsection{Other two-component fields}
We have addressed the question of the negative radiation pressure also for other models. Namely we have tested models of similar structure as in equation (\ref{lag2}):
\begin{equation}
{\cal L}=\frac{1}{2}\biggl(\phi_t^2-\phi_x^2-(\phi^2-1)^2\biggr)+\frac{1}{2}\left(\psi_t^2-\psi_x^2-m^2\psi^2\right)-V_{int}(\phi, \psi),
\end{equation}
where $V_{int}(\phi,\psi)$ is an interaction potential between fields $\phi$ and $\psi$ such that the interaction is weak far away from the kink, and large very close to the topological zero. In the present paper we have focused on potential:
\begin{equation}
 V^{(1)}_{int}(\phi,\psi) = \kappa \left(\phi^2-1\right)\psi.
\end{equation}
Another potential we have used had the following form:
\begin{equation}
 V^{(2)}_{int}(\phi,\psi) = \kappa \left(\phi^2-1\right)^2\psi.
\end{equation}
Although this potential defines theory with unbounded energy form below, the system is stable with respect to enough perturbation around the kink solution. To fix that inconvenience we could add some other potential which is negligible for small values of field and dominating for large values.

Other potentials (giving energy bounded from below) were:
\begin{equation}
 V^{(3)}_{int}(\phi,\psi) = \kappa \left(\phi^2-1\right)\psi^2.
\end{equation}
\begin{equation}
 V^{(4)}_{int}(\phi,\psi) = \kappa \left(\phi^2-1\right)^2\psi^4.
\end{equation}
For values $\kappa=0.2$, $A=0.2$, $\omega=2.5$ and $m=2.2$ we have performed numerical simulation and we have found that 
apart from $V^{(1)}_{int}$ only for potential $V^{(2)}_{int}$ the kink revealed the NRP. For the other two potentials the kinks were pushed by the radiation.

\subsubsection{A halfcompacton}
We have observed an interesting behaviour of a kink in a model of one interacting field with asymmetric potential described by lagrangian density:
\begin{equation}
{\cal L} = \frac{1}{2}\left(\phi_t^2-\phi_x^2-|\phi-1|(\phi+1)^2\right).
\end{equation} 
In this model the potential has two equal minima $\phi=\pm 1$, therefore topological defects are possible. Important difference from other well known models is that the filed of the kink approaches vacuum ($\phi=1$) at finite $x$ while on the other side it approaches vacuum exponentially. This type of object is usually referred to as a halfcompacton. Compactons are topological defects which differ from vacuum only in a finite segment \cite{Arodz}. They appear usually when the potential is V-shaped around vacua.

In this particular model the potential is smooth around one vacuum ($\phi=-1$) and sharp around the second one. Small perturbations around $\phi=-1$ can be described as field with mass $m^2=2$. Formally, the second derivative of the potential at the second vacuum is infinite and so is (formally) mass of small perturbation. Arbitrary small perturbations around the V-shaped potentials are nonlinear.

We have found that the kink is pushed when is hit with radiation coming from the vacuum $\phi=-1$. However it reveals NRP when hit with the wave is coming from the second vacuum $\phi=1$. The halfcompacton is always pushed towards the V-shaped vacuum. Again this process can be understood because of the mass difference ($m_1^2=2$ and formally $m_2^2=\infty$) of small excitation around two different vacua.

\section{Conclusions and discussion}
In the paper we have presented a model of two interacting scalar fields. 
Without interaction one of the fields ($\phi$) obeys $\phi^4$  and the second ($\psi$) Klein-Gordon equation with mass $m$.
This model reveals $\mathbb{Z}_2$ symmetry and topological defects, like kinks, are possible static solutions. 
We have shown that the kink, when exposed to scalar radiation of field $\psi$, can undergo a strange phenomenon of the negative radiation pressure. This happens when the mass parameter $m$ is larger than the mass of $\phi$ field which in our scaling is equal to $2$. Below that limit the kink is a subject to positive radiation pressure.
We have calculated analytically the force exerted on the kink by the radiation in the leading order both in amplitude of the wave and coupling constant. 
The force was proportional to product of squared amplitude and coupling constant.
The results were confirmed with numerical simulation of the full partial differential equation. The difference between analytical predictions and numerics were caused by sim\-pli\-fi\-ca\-tions used during calculations and preparation of initial conditions for numerical simulations. The descrapency was mostly a few percent.

The model can serve as an example of a different mechanism of negative radiation pressure than the one presented in our previous papers \cite{trom0,forgacs} in the case of $\phi^4$ model. The NRP in $\phi^4$ is a purely nonlinear effect and the force is proportional to the the fourth power of amplitude. The case of $\phi^4$ is a special one, due to the transparency of the kink, which is a rather rare feature. In other models kinks undergo usually positive radiation pressure. However, it is possible that there can be a critical amplitude above which the nonlinear effects, like NRP, are dominating. Another problem with the $\phi^4$ is that it is difficult to say how the kink would react to the non-monochromatic waves. As we have stated before the effect is nonlinear and no superposition is possible.

The NRP presented in the present paper has different properties. It  emerges in linear ap\-pro\-xi\-ma\-tion and therefore a superposition rule can be applied to it. This mechanism seems to be more generic than in the $\phi^4$ model. 
We did not need the transparency of the kink (in linear regime). 
The negative radiation pressure, although a very strange phenomenon, is probably more common than we had initially expected. 
We have changed the interaction potential between fields $\phi$ and $\psi$. Then we performed numerical simulation.
Some models revealed only positive radiation pressure. However, we have found a few models with NRP. 
For all of them there was a critical value of mass parameter $m_{cr}=2$ above which we observed NRP and below PRP. The observed acceleration was proportional to $A^2$. From whole that class the model presented in this paper was the simplest one. The simplicity allowed us to perform analytical calculations, which could be difficult in other models.

Our preliminary numerical simulations show that also vortices in Goldstone's and abelian Higgs model reveal NRP which is in its nature very similar to the one presented in this paper (see Appendix \ref{app:C}). It is also possible that it appears in other more complicated models, especially when there are many interacting fields. Possible candidates can be for instance models of kinks studied in \cite{Vachaspati}. 

Importance of the NRP is still an open question. However, it may be important in many physical processes where topological defects are present. For instance it may be important for stability of a system of topological defects. Any small excitation would radiate being a source of long-distance attractive interaction. This issue is clearly dimmension dependent. In 1+1 d the defects accelerating towards the source of radiation would eventually collide and bounce back or annihilate usually creating a long-living metastable object called oscillon \cite{Matzner}. These oscillons can also play important role in collisions with other objects. In higher dimensions the collisions are usually less frequent and the radiation pressure may be more important.

\appendix

\section{Noether's theorem}\label{sec:noether}
Our model possesses both time and spatial translational symmetry therefore we can define e\-ner\-gy-mo\-men\-tum tensor using the standard form:
\begin{equation}
   T^{\mu\nu}=\frac{\partial {\cal L}}{\partial\partial_\mu\phi}\partial_{\nu}\phi+\frac{\partial {\cal L}}{\partial\partial_\mu\psi}\partial_{\nu}\psi-{\cal L}\delta^{\mu\nu},
\end{equation} 
where as usual $T^{00}$ is energy density and $-T^{10}=-T^{01}$ is momentum density respectively.
The elements of this tensor are:
\begin{subequations}
\begin{equation}
T^{00} = {\cal E} = \frac{1}{2}\biggl(\phi_t^2+\phi_x^2+(\phi^2-1)^2\biggr)+\frac{1}{2}\left(\psi_t^2+\psi_x^2+m^2\psi^2\right)+
\kappa(\phi^2-1)\psi
\end{equation} 
\begin{equation}
   T^{01}=T^{10} = -{\cal P} = \phi_t\phi_x+\psi_t\psi_x
\end{equation} 
and 
\begin{equation}
    T^{11} = -\frac{1}{2}\biggl(\phi_t^2+\phi_x^2-(\phi^2-1)^2\biggr)-\frac{1}{2}\left(\psi_t^2+\psi_x^2-m^2\psi^2\right)+
\kappa(\phi^2-1)\psi.
\end{equation} 
\end{subequations}
Let us put a kink into a large segment $[-L,L]$. Using the Noether theorem we can find the rate of energy and momentum density flowing into this segment using only asymptotic forms of travelling waves.
The Noether theorem states that
\begin{equation}
   \partial_\mu T^{\mu\nu}= 0.
\end{equation} 
In particular, we can write
\begin{equation}
   \partial_t{\cal E} = \partial_0T^{00} = -\partial_xT^{10}
\end{equation} 
which after integration over the whole segment gives the energy carried by the wave into the segment
\begin{equation}\label{energy}
   \frac{\d E}{\d t}=\int_{-L}^{L}\!\!\dx\;\partial_xT^{10}=\phi_t\phi_x+\psi_t\psi_x\biggl.\biggr|_{-L}^L.
\end{equation} 
Moreover we also changed the sign to obtain the energy which flows into the segment and not is carried away by the wave.
In the very similar way we obtain the ratio of momentum flowing into the segment:
\begin{equation}\label{momentum}
\begin{aligned}
   \frac{\d P}{\d t} &= \int\!\!\dx\;\partial_0 T^{11}=\\
   &=-\frac{1}{2}\biggl(\phi_t^2+\phi_x^2-(\phi^2-1)^2\biggr)-\frac{1}{2}\left(\psi_t^2+\psi_x^2-m^2\psi^2\right)+
\kappa(\phi^2-1)\psi\biggl.\biggr|_{-L}^L.
\end{aligned}
\end{equation} 
Since we study interaction between a kink and a travelling waves, far away from the kink the filed has the following form (compare with equations (\ref{eq:travelling}, \ref{eq:travelling2})):
\begin{subequations}
\begin{equation}
   \phi(x,t) = \pm 1 + c_1\cos(\omega t+qx+\delta_1) + c_2\cos(\omega t+kx+\delta_2),
\end{equation}  

\begin{equation}
   \psi(x,t) = c_3\cos(\omega t+kx).
\end{equation} 
\end{subequations}
Substitution of the above into (\ref{energy}) and (\ref{momentum}) and averaging over time gives:
\begin{subequations}
   \begin{equation}
      \left\langle\frac{\d E}{\d t}\right\rangle_T=\frac{1}{2}\omega q c_1^2 + \frac{1}{2}\omega k c_2^2 + \frac{1}{2}\omega k c_3^2\biggl.\biggr|_{-L}^L.
   \end{equation} 
As one can see from (\ref{eq:travelling}) the amplitude of second term in $\phi(x,t)$ is the same on both sides of the kink and therefore they cancell each other out. In the similar way we obtain
   \begin{equation}
      \left\langle\frac{\d P}{\d t}\right\rangle_T=-\frac{1}{2}q^2 c_1^2 - \frac{1}{2} k^2 c_3^2-\frac{1}{2}k^2c_2^2+\kappa c_2c_3\biggl.\biggr|_{-L}^L.
   \end{equation} 
\end{subequations}
Note that form equations (\ref{eq:travelling}, \ref{eq:travelling2}) and following discusion we can see that terms with $c_2$ either have the same values at both ends of a segment or they differ by at least of order $\Or(\kappa^4)$ therefore they do not give contribution to equations (\ref{eq:energy}) and (\ref{eq:force}).

\section{Asymptotic form of linear solution}
In our perturbation scheme we assumed that the coupling constant is small. We have obtained the solutions which asymptotically are waves with wave number equal to $k=\sqrt{\omega^2-m^2}$ and $q=\sqrt{\omega^2-4}$. It is easy to show that in the linear regime, far away from the kink, the wave numbers are different. This difference should appear at some point further in our perturbation scheme.
Let us seek for solutions in the form
 \begin{equation}
   \phi(x,t)=\phi_0(x)+A\e^{\i \omega t}\xi(x),\;\;\phi=\psi_0(x)+A\e^{\i \omega t}\chi(x)
\end{equation} 
assuming that $A$ is small. After substituting the above into the equations of motion we obtain:
\begin{equation}
   \begin{bmatrix}
{\displaystyle-\frac{\d^2}{\dx^2}+2(3\phi_0^2-1)+2\kappa\psi_0}&2\kappa\phi_0\\
      2\kappa\phi_0&{\displaystyle-\frac{\d^2}{\dx^2}+m^2}\\
   \end{bmatrix}
   \begin{bmatrix}
      \xi\\\chi
   \end{bmatrix}
=\omega^2\begin{bmatrix}
      \xi\\\chi
   \end{bmatrix}
\end{equation} 
Far away from the kink, for $x\rightarrow\pm\infty$ $\psi_0\rightarrow0$ and $\phi\rightarrow \pm 1$. In this limit the equations take the form:

\begin{equation}
 \begin{bmatrix}
{\displaystyle-\frac{\d^2}{\dx^2}+4}&\pm2\kappa\\
      \pm2\kappa&{\displaystyle-\frac{\d^2}{\dx^2}+m^2}\\
   \end{bmatrix}
   \begin{bmatrix}
      \xi\\\chi
   \end{bmatrix}
=\omega^2\begin{bmatrix}
      \xi\\\chi
      \end{bmatrix}
\end{equation}
The eigensolutions are:
\begin{equation}
   \begin{bmatrix}
      \xi\\\chi
      \end{bmatrix}=
      \begin{bmatrix}
      \displaystyle \pm\frac{2\kappa}{4-\omega^2+k_{1,2}^2}\\1
      \end{bmatrix}\e^{\i k_{1,2}x},
\end{equation} 
where
\begin{equation}
   k^2_{1,2} = -\frac{m^2+4-2\omega^2\pm\sqrt{(m^2-4)^2+16\kappa^2}}{2}.
\end{equation} 
For small values of $\kappa$
\begin{equation}
  k_1^2={\omega}^{2}-{m}^{2}-4{\frac {{\kappa}^{2}}{{m}^{2}-4}}+\Or(\kappa^4)
\end{equation} 
and 
\begin{equation}
  k_2^2={\omega}^{2}-4+4{\frac {{\kappa}^{2}}{{m}^{2}-4}}+\Or(\kappa^4).
\end{equation} 
The above results could be used as  a test of how accurate our expansion in coupling constant is. 
One can see that the first correction to the wave numbers is of order $\Or(\kappa^2)$.
Note that for $m$ not much different from $2$ our expansion fails.
\section{Application to vortices}\label{app:C}
In this appendix we shortly discuss application of the mechanism described in the present paper to more physical, yet more complicated models.
\subsection{Goldstone's model}
Let us consider a complex scalar field described by the lagrangian:
\begin{equation}
{\cal L} = \partial_\mu\phi\partial^\mu\phi^*-\left(\phi^*\phi-1\right)^2.
\end{equation} 
This model reveals global $U(1)$ symmetry. Vacuum manifold is a circle $|\phi|=1$. 
Static non-trivial solutions with cylindrical symmetry (vortices) are therefore possible.
Let us limit our conciderations to cylindrical symmetry.
The vortex solution has a form
\begin{equation}
   \phi(r,\theta) = f(r)\e^{\i N\theta},
\end{equation} 
where $N$ is a winding number (in this model only vortices with $N=1$ are energetically stable).
Perturbations, both around the vacuum state and a vortex, are of two types. So called Goldstone's mode is an excitation along the vacuum manifold (phase changing) and has mass equal to 0 ($m_g=0$).
The second perturbation excites the field perpendicularly to the vacuum manifold (changes field's amplitude, therefore we will call this excitation an amplitude mode). This mode has mass equal to 2 ($m_a=2$).

Our preliminary numerical simulations show that vortices are pushed by planar wave generated from Goldstone's mode.  Amplitude wave exerts negative radiation pressure. The motion of the vortex is more complicated because vortices in this model are not well localized and have divergent energy.
The phenomenon of the negative radiation pressure in case of vortices can be understood qualitatively using the same arguments as in the present paper.
Because of the mass difference the amplitude mode creates a surplus of momentum behind the vortex.
Numerical results along with analytical calculations in this case will be presented in a separate paper \cite{future1}.
\subsection{Abelian Higgs model}
Another very often studied model is abelian Higgs model for which the lagrangian has a form:
\begin{equation}
{\cal L} = D_{\mu}\phi(D^\mu\phi)^*-\frac{\lambda}{2}\left(\phi^*\phi-\eta^2\right)^2-\frac{1}{4}F_{\mu\nu}F^{\mu\nu},
\end{equation} 
where $D_\mu=\partial_\mu-\i e A_\mu$ and $F_{\mu\nu}=\partial_\mu A_\nu-\partial_\nu A_\mu$ is a Faraday tensor.
Vortices are possible static solution due to the (local) $U(1)$ symmetry. Vortices in this model are more localized than those in Goldstone's mode.
Again, our preliminary numerical simmulation show, that they are a subject to the NRP when hit with amplitude wave. Moreover the NRP also appears when the vortex is exposed to vector-field wave with appropriate polarization. We belive that this phenomenon has exactly the same mechanism as the one presented in the present paper: the mass difference between those fields (amplitude mode, phase mode and vector field) is responsible for this effect.

\section*{ACKNOWLEDGEMENTS}
We wish to thank H.~Arodz, P.~Forgacs ans A.~Lukacs for many hours of interesting conversations and many helpful remarks.


\begin{thebibliography}{16}
\bibitem{trom0} T.~Roma\'nczukiewicz, {\it Acta Phys.~Pol.} {\bf  B35}, 523 (2004).
\bibitem{forgacs} P.~Forg\'acs, \'A.~Luk\'acs, T.~Roma\'nczukiewicz, {\it Phys.~Rev.}{\bf  D77}, 125012 (2008).
\bibitem{future1} P.~Forg\'acs, J.~Karkowski, \'A.~Luk\'acs, T.~Roma\'nczukiewicz, {\it in preparation},  (2008).
\bibitem{Vachaspati} T.~Vachaspati {\it Phys.~Rev.} {\bf D64}  105023 (2001).
\bibitem{Matzner} P.~Anninos,  S.~Oliveira, R.~A.~Matzner {\it Phys.~Rev.} {\bf D44} 1147 (1991)
\bibitem{Arodz} H.~Arod\'z {\it Acta Phys.~Polon.} {\bf B33}  1241 (2002)
\end{thebibliography}
\end{document}